\documentstyle[sprocl,epsfig]{article}

\def\be{\begin{equation}}
\def\ee{\end{equation}}
\def\bea{\begin{eqnarray}}
\def\eea{\end{eqnarray}}

\begin{document}

\title{Non-Gaussian spectra and the search for cosmic strings}

\author{Jo\~ao Magueijo and Alex Lewin}

\address{The Blackett Laboratory, Imperial College,
Prince Consort Road, London SW7 2BZ, UK}


\maketitle\abstracts{We present a new tool for relating theory 
and experiment suited for non-Gaussian theories: non-Gaussian 
spectra. It does for non-Gaussian theories what the angular power 
spectrum $C_\ell$ does for Gaussian theories. We then show how 
previous studies of cosmic strings have  over rated
their non-Gaussian signature. More realistic maps are not visually
stringy. However non-Gaussian spectra will accuse their stringiness.
We finally summarise the steps of an undergoing experimental project
aiming at searching for cosmic strings by means of this technique. 
}
  
\section{Theory and experiment in CMB physics}
In this review we present a new tool for making contact between theory
and experiment in cosmic microwave background (CMB) physics \cite{fermag}. 
The
central tool which has played this role in the past is the angular 
power spectrum (known as $C_\ell$). If the underlying theory is Gaussian
it is known that the power spectrum $C_\ell$ encodes the totality of
the predictions made by the theory. It then makes sense to direct all 
experimental effort towards its measurement. Also the power spectrum 
represents the data-reduction end-point should we firmly believe
in the Gaussianity of the underlying Universe. If we start with an all
sky map with $10^6$ pixels one could reduce them to roughly $10^3$
values for $C_\ell$. For Gaussian theories all the science is
in these $10^3$ numbers. The map itself is redundant.
If we further believe not only that the Universe is Gaussian,
but also that inflation is the truth, then these $C_\ell$ are only dependent 
on about $10$ parameters. Therefore we may in fact carry this 
reduction further to about $10$ quantities.

If the underlying theory is non-Gaussian, however, 
the power spectrum is not the end of the story. In this paper we define  
a new set of spectra, which we will label non-Gaussian
spectra, and which carry predictive power in generic 
non-Gaussian theories. We will argue that non-Gaussian spectra could
in fact be the best arena for connecting theory and experiment
in motivated non-Gaussian theories such as cosmic string
scenarios. Although the estimation of
these spectra complicates data-reduction and data-analysis considerably,
non-Gaussian spectra are essential for bringing out the full predictive 
power of non-Gaussian theories. They may also help to quantify  in which sense topological
defect theories are non-Gaussian. This becomes particularly relevant
if one allows for the full complication of theories like cosmic strings
to be included in the consideration of their non-Gaussian signals.

It should be stated from the start that if the underlying theory is indeed
Gaussian, non-Gaussian spectra are perfectly useless. One may not,
for instance, improve estimates of cosmological parameters in 
inflationary scenarios through their measurement. Prejudice is however
a nasty element in science, and one should think of
non-Gaussianity as an issue in its own right, regardless
of any theoretical dogma. Most data-analysis methods, in 
particular, rely blindly on Gaussianity. They could well break down
miserably should the data prove to be non-Gaussian in the first place.

\section{A taste of non-Gaussian spectra}
The idea of non-Gaussian spectra is to supplement the angular power 
spectrum with a set of quantities whose number may never exceed the
number of information degrees of freedom one starts from. If $N$ independent
pixels have been measured then transforming them into a larger number
of quantities obviously introduces redundancy. On the other hand computing
a small number of quantities, such as the skewness and kurtosis, 
or in general submitting the data to Gaussianity tests,
may leave room for all sorts of perverse and less perverse
non-Gaussianity to pass unnoticed. 
In general our philosophy is to factor
out from the data only rotational degrees of freedom 
(how we oriented the axes). There are 3 degrees of freedom in rotations. 
There should then be $N-3$ quantities with
which to compute the power spectrum, plus the non-Gaussian spectra.

Let us consider the spherical harmonic expansion of an all-sky map:
\begin{equation}
\frac{\Delta T(\bf{n})}{T}   
= \sum_{\ell=0}^{\infty} \sum_{m=-\ell}^{\ell}
a^{\ell}_ m Y^{\ell}_m (\bf{n})
\end{equation}
Consider also the situation where the field is very small so that
instead one may expand in Fourier modes:
\begin{equation}\label{fourier}
  \frac{\Delta T({\bf x})}{T}
={\int {d{\bf k}\over 2\pi}a({\bf k})e^{i{\bf k}\cdot{\bf x}}}
\end{equation} 
There is some correspondence between $a^\ell_m$ and $a({\bf k})$.
To some extent the moduli $|{\bf k}|$ correspond to $\ell$,
and act as the scale of the modes. On the other hand the $m$ is
a bit like the direction of the vector ${\bf k}$ and labels something
like the direction of the mode. One must remember that these modes are 
complex, and that therefore have phases as well as moduli (which are
the only components entering measures of power).

If for simplicity one concentrates on the Fourier space picture,
then computing the power spectrum may be seen as the result of the following
algorithm. One divides Fourier space into rings with $\Delta k=1$,
and then averages the square of the moduli of modes lying in each
of these rings. The average $C_\ell$ (or $C(k)$) represents a measure
of the amplitude (or the power) of the fluctuations on the scale $\ell$ 
(or $|{\bf k}|$):
\be
C(k)={\langle |a({\bf k})|^2\rangle}
\ee
Within this picture one can think of a natural way of extending
this construction as a set of transverse or ring spectra. These
represent a measure of how the power is distributed across each ring
in angle. Such a measure of angular distribution of power would naturally
represent the shape of the fluctuations. However one may now talk about
shape on a given scale, something which is far from visually intuitive.

\begin{figure}
\centerline{\epsfig{file=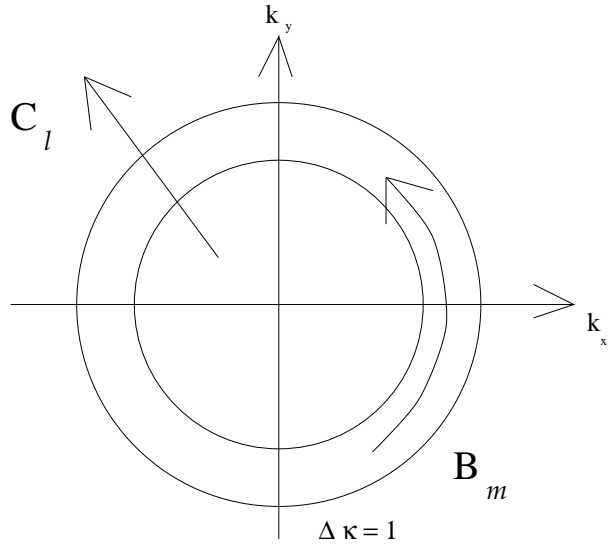,angle=-90,width=8cm}}
\caption{A schematic picture of power spectra and non-Gaussian
spectra. Dividing Fourier space into rings with $\Delta k=1$
one obtains the power spectrum $C_\ell$ by averaging the power
on each ring. Non-Gaussian spectra $B_m$ are a measure of how
the power is distributed in angle, in each ring. Phase spectra
and inter-ring spectra complement this, so-called, shape spectra.}
\label{fig1}
\end{figure}

On top of this one can consider phase spectra. Phases transform
under translations and therefore represent the localisation of
the fluctuations.  Again we may define localisation as a function
of scale, something rather abstract. Finally one can consider a
radial spectrum of correlations between adjacent rings. 
Inter-ring correlators are the crucial aspect of non-Gaussianity which allows 
for shape and localisation on the various scales to be transmuted
into structures  which we can recognise visually. Indeed one needs
the constructive interference between all the scales so that something
as abstract as shape and place on a given scale becomes, say, the picture
of an elephant.

The advantage of such abstract definitions of shape and place is precisely
in that they do not rely on visual recognition. As we will see soon
very rarely life is so kind to us so as to provide us with evident
visual non-Gaussianity. A more abstract, but still comprehensive,
method for describing shape and place is therefore necessary.

\section{Realistic stringy skies}

\begin{figure}
\centerline{\epsfig{file=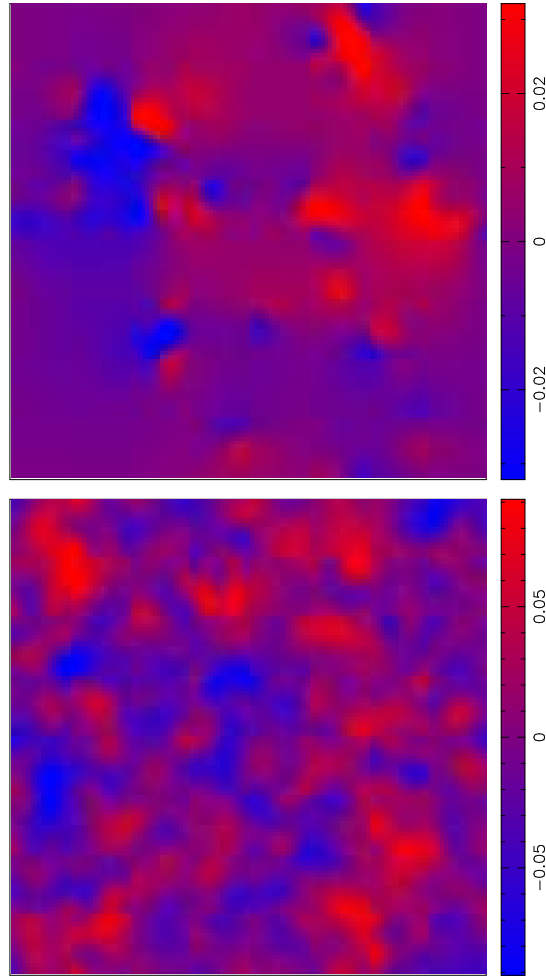,angle=0,width=10cm}}
\caption{The Kaiser Stebbins effect from cosmic strings (top)
if the glow out of the last scattering surface were homogeneous.
Adding on the perturbations induced by strings before last 
scattering leaves what looks like a Gaussian map (bottom).}
\label{fig2}
\end{figure}

The formalism we have just outlined acts as comprehensive formalism
for encoding non-Gaussianity, which is an advantage over Gaussianity
tests. Also, unlike the $n$-point correlation function, it does not
contain redundant information. However it lives naturally in Fourier 
space. This contradicts a well established dogma: non-Gaussianity
is obvious in real space, but gets diluted in Fourier space.
According to this argument a Fourier mode is the result of
the addition of many possibly non-Gaussian temperatures:
\begin{equation}\label{fourier2}
  a({\bf k})
={\int {d{\bf k}\over 2\pi}\frac{\Delta T({\bf x})}{T} 
e^{-i{\bf k}\cdot{\bf x}}}
\end{equation} 
The central limit theorem then tells us that even if the 
$\frac{\Delta T({\bf x})}{T}$ are very non-Gaussian the $a({\bf k})$
will be approximately Gaussian. As an example we may consider
the Kaiser-Stebbins effect from cosmic strings (see Fig.~\ref{fig2},
top figure). Distinctive stringy discontinuities may be recognised
in the top map. The Fourier transform of such a map would be a bit
of a mess, a mess that is not particularly non-Gaussian.

We will counter this argument using precisely the example of cosmic string 
maps. These have been oversimplified in the past, in a way that 
over rates their non-Gaussinity. A more realistic examination
of stringy skies does not comply with the cubist microwave sky often 
attributed to cosmic strings. To begin with realistic cosmic strings 
are rather contorted objects. As the top figure in Fig.~\ref{fig2}
shows one does see jumps on stringy skies but these have a rather
limited coherence length.

More important than this is the recognition that even the top
picture in Fig.~\ref{fig2} is a simplification. It assumes that
the glow of photons coming out of the last scattering surface 
is perfectly homogeneous. However one must remember that there
were  strings before last scattering. These also caused
fluctuations in the cosmic radiation. After last scattering
radiation is made up of free photons that fly past the moving strings
and ``take a picture'' of the string network. Before last
scattering the photons behaved more like a fluid, which was slashed
by the strings, but was also subject to its own pressure. A complete
mess of waves may be expected, rather than a neat picture of
the string network. Indeed the only available calculation of
string perturbations before last scattering shows rather Gaussian
looking fluctuations \cite{neil}.

\begin{figure}
\centerline{\epsfig{file=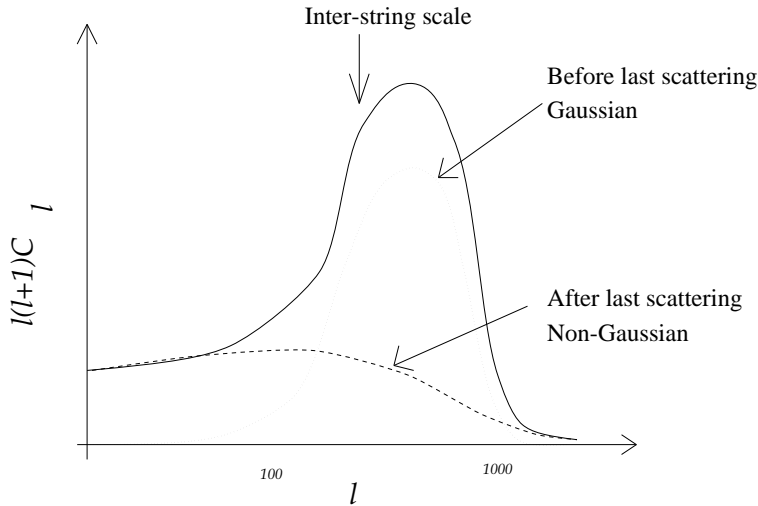,angle=-90,width=10cm}}
\caption{Schematic representation of the power spectrum of the
Gaussian fluctuations induced by cosmic strings before last scattering
and the non-Gaussian fluctuations induced after last scattering by
the Kaiser-Stebbins effect. We also indicate the scale of the 
inter-string separation at last scattering, which is $\theta\approx
20'$ corresponding to $\ell\approx 60^\circ/\theta\approx 200$.}
\label{fig3}
\end{figure}
One may plot the power spectrum of the fluctuations
induced by cosmic strings schematically as in Fig.~\ref{fig3}.
The $C_\ell$ may be seen as the result of two contributing 
components. One corresponds to fluctuations induced by strings
after last scattering, which gives the beautiful Kaiser-Stebbins
effect  one usually pictures. The other corresponds to the fluctuations 
induced before last scattering, which should
be roughly Gaussian. The non-Gaussian component dominates on large
scales, presenting a slightly tilted spectrum \cite{allen}, then falls off
as a power law \cite{mark}. The Gaussian component is negligible at large
scales (white noise, rather than quasi-scale invariant), then rises
into a single Doppler peak of height as yet unknown, but roughly
placed at $\ell\approx 400-600$ \cite{andy}. It then falls off exponentially,
due to Silk damping 
\footnote{We believe that the point made in \cite{rich} in purely semantic,
concerning what to call after and before last scattering. Whatever the
before/after definition is, the power in the Gaussian component 
should always fall off exponentially}. 

\begin{figure}
\centerline{\epsfig{file=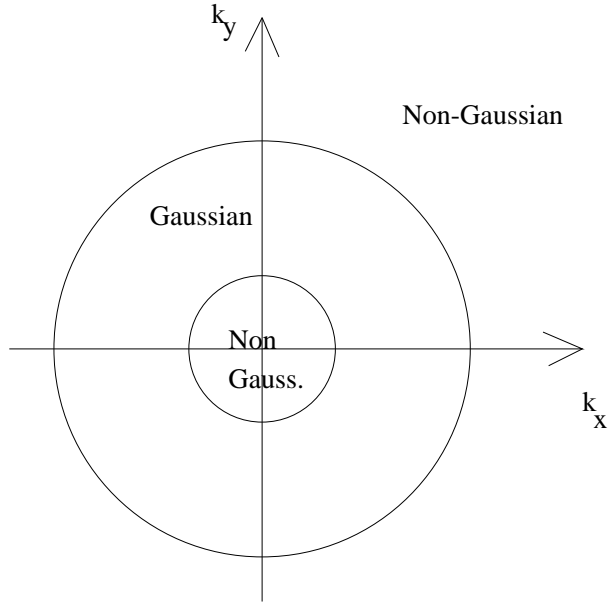,angle=-90,width=8cm}}
\caption{The Fourier space structure of realistic string
skies.}
\label{fig4}
\end{figure}
If the Gaussian component were not there all one would need would be an 
experiment with a resolution better than the inter-string separation
at last scattering, and the Kaiser-Stebbins effect would emerge
in all its glory. However at the inter-string separation scale
the power of the fluctuations is dominated by the Gaussian component.
Still not all is lost. Because the Gaussian component falls off
as an exponential, whereas the non-Gaussian component falls off
like a power law, at very high $\ell$ the fluctuations should
be dominated by the non-Gaussian component again. All in all
the Fourier space image of a stringy sky may be divided into mainly
Gaussian and mainly non-Gaussian bands, as depicted in Fig.~\ref{fig4}.
It looks as if realistic string maps lead to the conclusion
that one needs much higher resolution to detect strings than
previously thought, and that their detection will probably be clearer 
in the Fourier domain.

We are now in position to reverse the central limit theorem
argument so often used against non-Gaussianity analysis in the
Fourier domain. The real space temperature is now seen
as the sum of many Fourier space components, some of which
are strongly non-Gaussian, and some others nearly Gaussian:
\begin{equation}
  \frac{\Delta T({\bf x})}{T}
={\int {d{\bf k}\over 2\pi}a({\bf k})e^{i{\bf k}\cdot{\bf x}}}
\end{equation}
The central limit theorem can now, even more than before, predict
that the real space imagine will be nearly Gaussian. The bottom
figure in Fig.~\ref{fig2} shows that indeed the realistic Kaiser-Stebbins
effect does not lead to any structures which any conceivable
Human could visually recognise.

Clearly we have a situation where the Gaussian and non-Gaussian 
components are naturally separated in Fourier space. The non-Gaussian
signal in Fourier space will never be a visual signal. However
it should be now clear that if I find an algebraic way to detect
stringiness in Fourier space, this method will be robust against
the addition of Gaussian signal or even Gaussian noise, as long
as a non-Gaussian band in Fourier space survives these additions.

Non-Gaussian spectra provide such an algebraic tool. They define shape
on a given scale. They should therefore allow the detection of
stringy shapes on the outer non-Gaussian band where the stringy
Kaiser-Stebbins effect dominates the signal.

\section{Non-Gaussian spectra}
A more mathematical definition of non-Gaussian spectra will now be given.
Consider a ring of the Fourier space where $N_k$ independent
complex modes $a({\bf k}_i)=\Re [a({\bf k}_i)] 
+i\Im [ a({\bf k}_i)]$ live.
In Gaussian theories these are distributed as
\begin{eqnarray}
  F(\Re[a({\bf k}_i)],\Im[a({\bf k}_i)])=
      {1\over (2\pi\sigma^2)^{N_k/2}}\times\nonumber \\
        \exp{-\left(
          {1\over 2\sigma_k^2}\sum_{i=1}^{m_k}(\Re^2[a({\bf k}_i)]+
          \Im^2[a({\bf k}_i)])\right)}
\end{eqnarray}
where $m_k=N_k/2$.
First separate the $N_k$ complex modes into $m_k$ moduli $\rho_i$
and $m_k$ phases $\phi_i$
\begin{eqnarray}
  \Re[a({\bf k}_i)]&=&\rho_i\cos{\phi_i}\nonumber\\
  \Im[a({\bf k}_i)]&=&\rho_i\sin{\phi_i}
\end{eqnarray}
The Jacobian of this transformation is 
\begin{equation}\label{jac1}
  \left| {\partial (\Re[a({\bf k}_i)],\Im[a({\bf k}_i)])
      \over \partial(\rho_i,\phi_i)}\right|=\prod_{i=1}^{m_k}\rho_i
\end{equation}
The  $\{\rho_i\}$ may  be seen as 
Cartesian coordinates which we transform into polar coordinates.
These consist of a radius $r$ plus $m_k-1$ angles $\tilde\theta_i$
given by
\begin{equation}\label{pol}
  \rho_i=r\cos{\tilde\theta_i}\prod_{j=0}^{i-1}\sin{\tilde\theta_j}
\end{equation}
with $\sin{\tilde\theta_0}=\cos{\tilde\theta_{m_k}}=1$.
In terms of these variables the radius is related to the
angular power spectrum by $C(k)=r^2/(2m_k)$. In general the first 
$m_k-2$ angles $\tilde\theta_i$ vary between $0$ and $\pi$ and the 
last angle varies between 0 and $2\pi$.
However because all $\rho_i$ are positive all angles are in $(0,\pi/2)$.
The Jacobian of this transformation is 
\begin{equation}\label{jac2}
  \left| {\partial(\rho_1,\cdots,\rho_{m_k})
      \over \partial(r,\tilde\theta_1,\cdots,\tilde\theta_{m_k-1}}\right|=
     r^{m_k-1}\prod_{i=2}^{m_k-1}\sin^{m_k-i}{\tilde\theta_{i-1}}
\end{equation}
Polar coordinates in $m_k$ dimensions may be understood as the iteration
of the following rule:
\begin{eqnarray}
  \rho_i&=&r_i\cos{\tilde\theta_i}\nonumber \\
  r_{i-1}&=&r_i\sin {\tilde\theta_i}
\end{eqnarray}
in which $r_i$ is the radius of the shade $m_k-i+1$ dimensional sphere
obtained by keeping fixed all $\rho_j$ for $j=1,\cdots,i-1$:
\begin{equation}
  r_i={\sqrt{\rho_i^2+\rho_{i+1}^2+\cdots+\rho_{m_k}^2}}
\end{equation}
One may easily see that this is how 3D polars work, and also that
the transform (\ref{pol}) follows this rule.
Hence one may invert the transform (\ref{pol}) with
\begin{equation}
  \tilde\theta_i=\arccos{\rho_i\over {\sqrt{\rho_i^2+\rho_{i+1}^2+
        \cdots+\rho_{m_k}^2}}}
\end{equation}
for $i=1,\cdots,m_k-1$. 

The Jacobian of the transformation
from $(\Re[a({\bf k}_i)],\Im[a({\bf k}_i)])$ 
to $\{r,\tilde\theta_i,\phi_i\}$ is
just the product of (\ref{jac1}) and (\ref{jac2}).
Hence for a Gaussian theory one has the distribution
\begin{equation}
  F(r,\tilde\theta_i,\phi_i)={
    {r^{N_k-1}\exp{-\left(r^2\over 2\sigma_k^2\right)}}
      \over (2\pi\sigma^2)^{N_k/ 2}}
          \prod_{i=1}^{m_k-1}\cos{\tilde\theta_i}
          (\sin{\tilde\theta_i})^{N_k-2i-1}
\end{equation}
In order to define $\tilde\theta_i$ variables 
which are uniformly distributed in 
Gaussian theories one may finally perform the transformation on each
$\tilde\theta_i$:
\begin{equation}
  {\theta_i}=\sin^{N_k-2i}(\tilde\theta_i)
\end{equation}
so that for Gaussian theories one has:
\begin{equation}
  F(r,{\theta}_i,\phi_i)={r^{N_k-1}e^{-r^2/(2\sigma_k^2)}
    \over 2^{m_k-1}(m_k-1)!}\times 1 \times\prod_{i=1}^{m_k}{1\over
    2\pi}
\end{equation}
The factorization chosen shows that all new variables are independent
random variables for Gaussian theories. $r^2$ has a $\chi^2_{N_k}$
distribution,
the ``shape'' variables $\theta_i$ are uniformly distributed
in $(0,1)$, and the phases $\phi_i$ are uniformly distributed in $(0,2\pi)$.

The  variables $\theta_i$ define a non-Gaussian shape spectrum,
the {\it ring spectrum}. 
They may be computed from ring moduli $\rho_i$  simply by
\begin{equation}
  {\theta}_i={\left(\rho_{i+1}^2+\cdots +\rho_{m_k}^2
      \over \rho_i^2\cdots +\rho_{m_k}^2\right)}^{m_k-i}
\end{equation}
They describe how shapeful the perturbations are. 
If the perturbations are stringy then
the maximal moduli will be much larger than the minimal moduli.
If the perturbations are circular, then all moduli will be roughly
the same. This favours some combinations of angles, which are
otherwise uniformly distributed. In general any shapeful picture
defines a line on the ring spectrum $\theta_i$.
A non-Gaussian theory ought to define a set of probable smooth
ring spectra peaking along a ridge of typical shapes.

We can now construct an invariant for each adjacent pair of
rings, solely out of the moduli. If we order the $\rho_i$ for each
ring, we can identify the maximum moduli. Each of these moduli
will have a specific direction in Fourier space; let 
 ${\bf k}_{max}$
and ${\bf k}^{'}_{max}$ be the directions where the maximal moduli
 are achieved.
The angle
\begin{equation}
  \psi(k,k')={1\over \pi}{\rm ang}({\bf k}_{max},{\bf k}^{'}_{max})
\end{equation}
will then produce an inter-ring correlator for the moduli, the
{\it inter-ring spectra}. This 
is uniformly distributed in Gaussian theories in $(-1,1)$. It gives
us information on how connected the distribution of power is between
the different scales. 

We have therefore defined a transformation from the original modes
into a set of variables $\{r,\theta,\phi,\psi\}$. The non-Gaussian
spectra thus defined have a  particularly simple distribution
for Gaussian theories. The fact that this distribution does not
have a peak shows clearly that we cannot use non-Gaussian spectra
to find out anything about the parameters of a Gaussian theory
(eg. cosmological parameters in inflation).
We shall call perturbations for which the phases are not uniformly
distributed localized perturbations. This is because if perturbations
are made up of lumps statistically distributed but with well defined 
positions then the phases will appear highly correlated. We shall
call perturbations for which the ring spectra are not
uniformly distributed shapeful perturbations. We will identify later
the combinations of angles which measure stringy or spherical shape of the
perturbations. This distinction is interesting as it is in principle
possible for fluctuations to be localized but shapeless, or more 
surprisingly, to be shapeful but not localized. Finally we shall call 
perturbations for which the inter-ring spectra are not uniformly
distributed, connected perturbations. This turns out to be one of
the key features of stringy perturbations. These three definitions
allow us to consider structure in various layers. White noise
is the most structureless type of perturbation. Gaussian fluctuations
allow for modulation, that is a non trivial power spectrum $C(k)$,
but their structure stops there.
Shape, localization, and connectedness constitute the three next
levels of structure one might add on. Standard visual structure
is contained within these definitions, but they allow for more
abstract levels of structure. 

\section{The meaning of non-Gaussian spectra}
\begin{figure}
\centerline{\epsfig{file=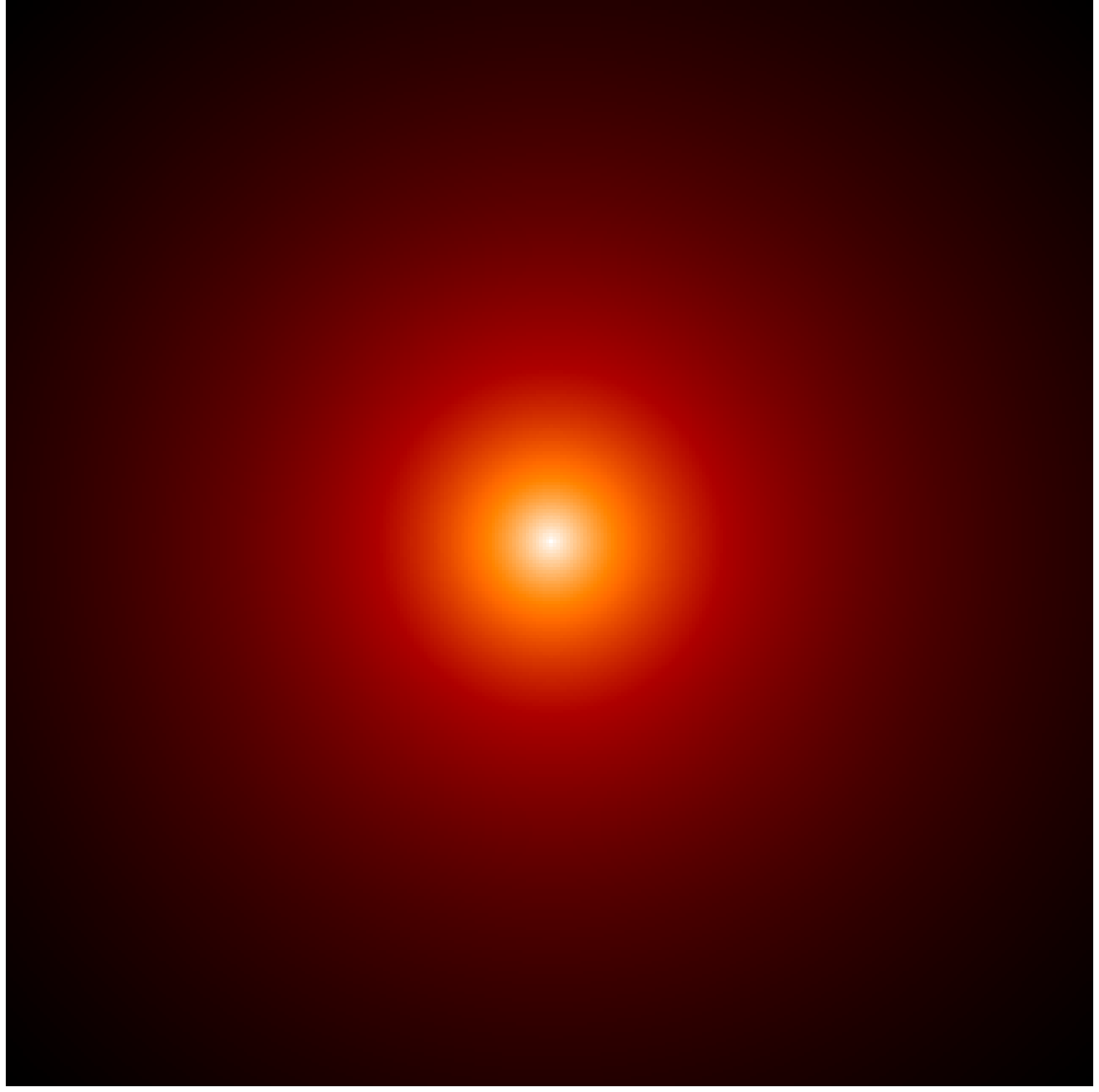,angle=0,width=6cm}
\epsfig{file=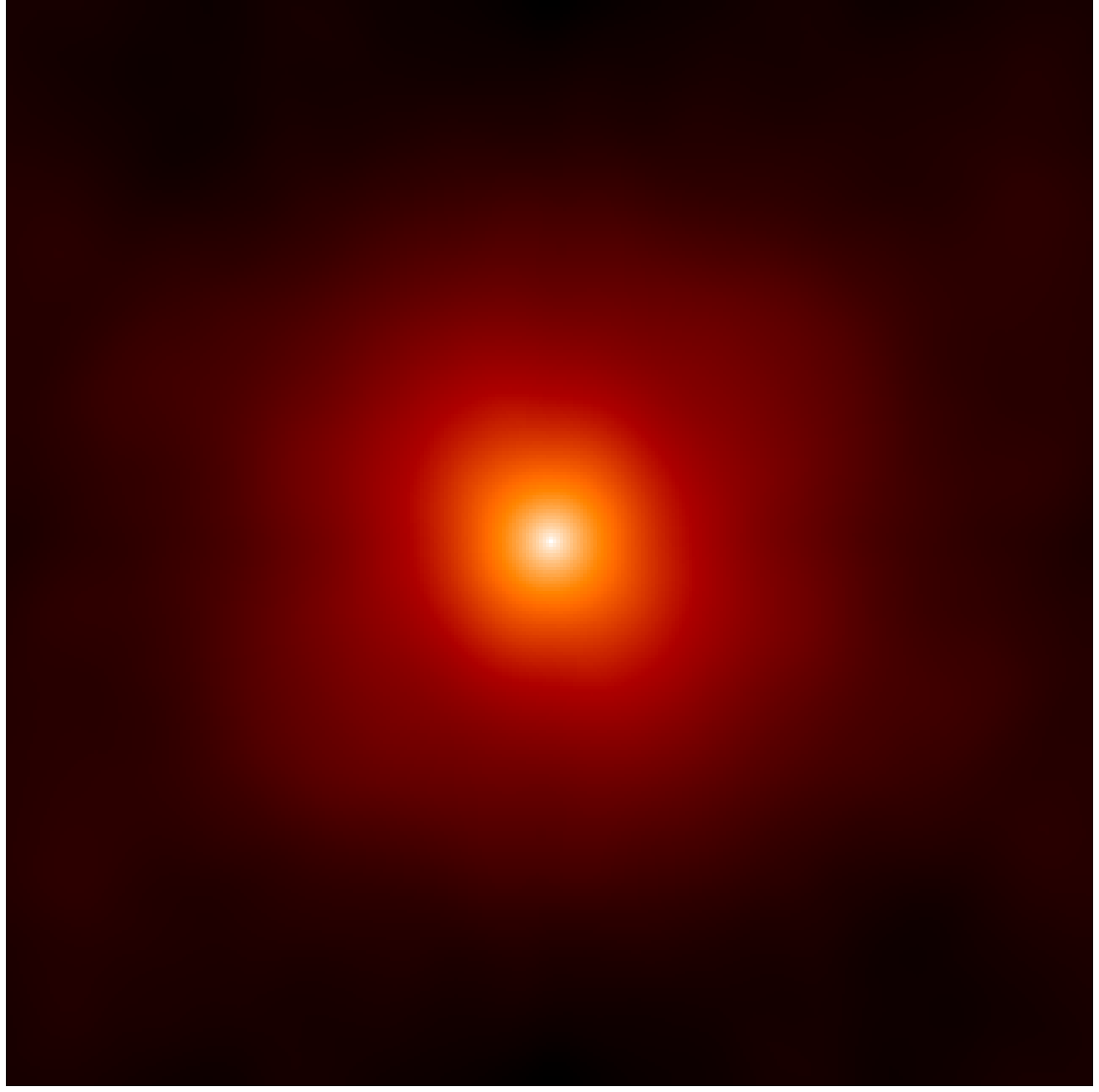,angle=0,width=6cm}}
\centerline{\epsfig{file=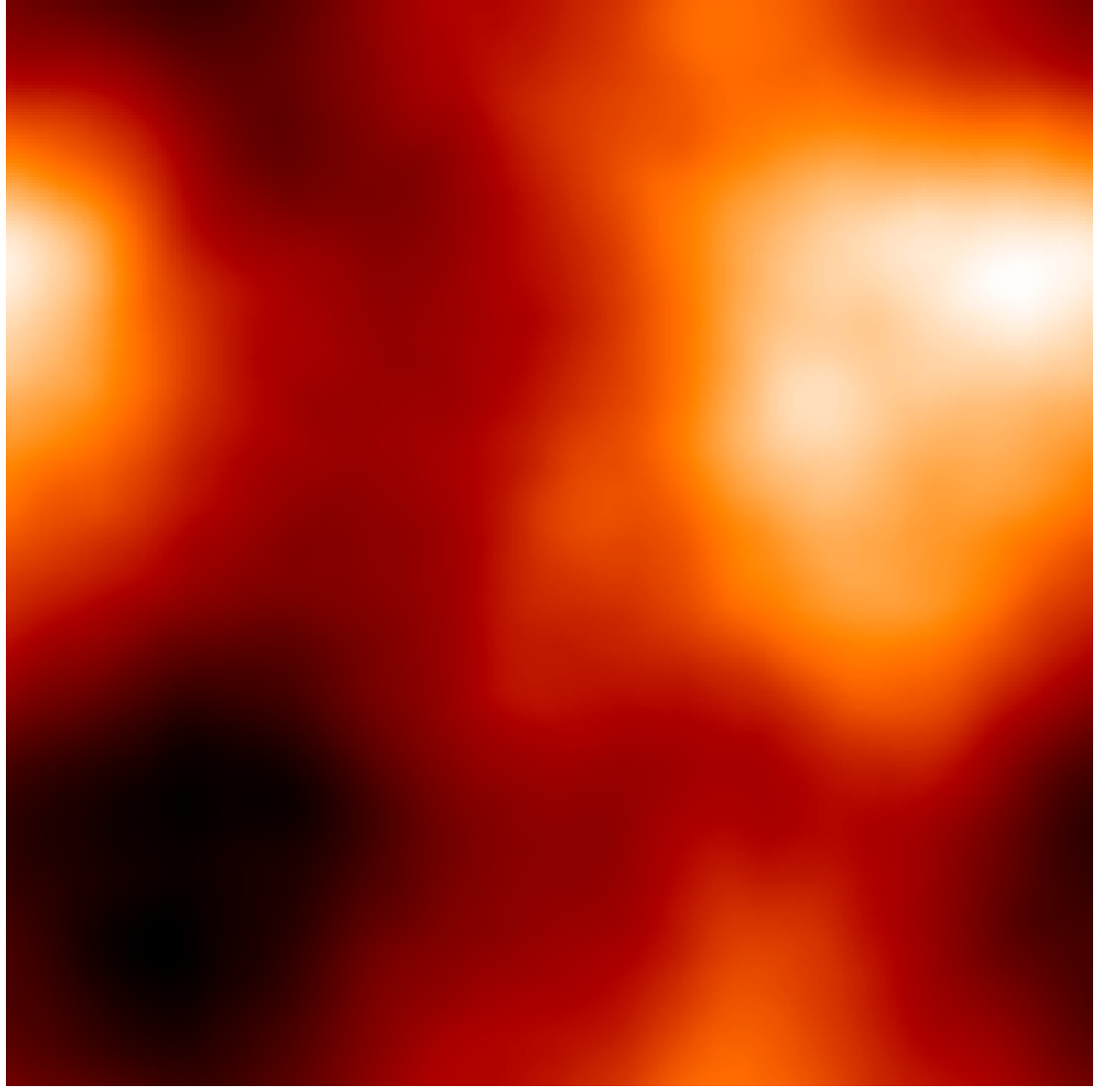,angle=0,width=6cm}
\epsfig{file=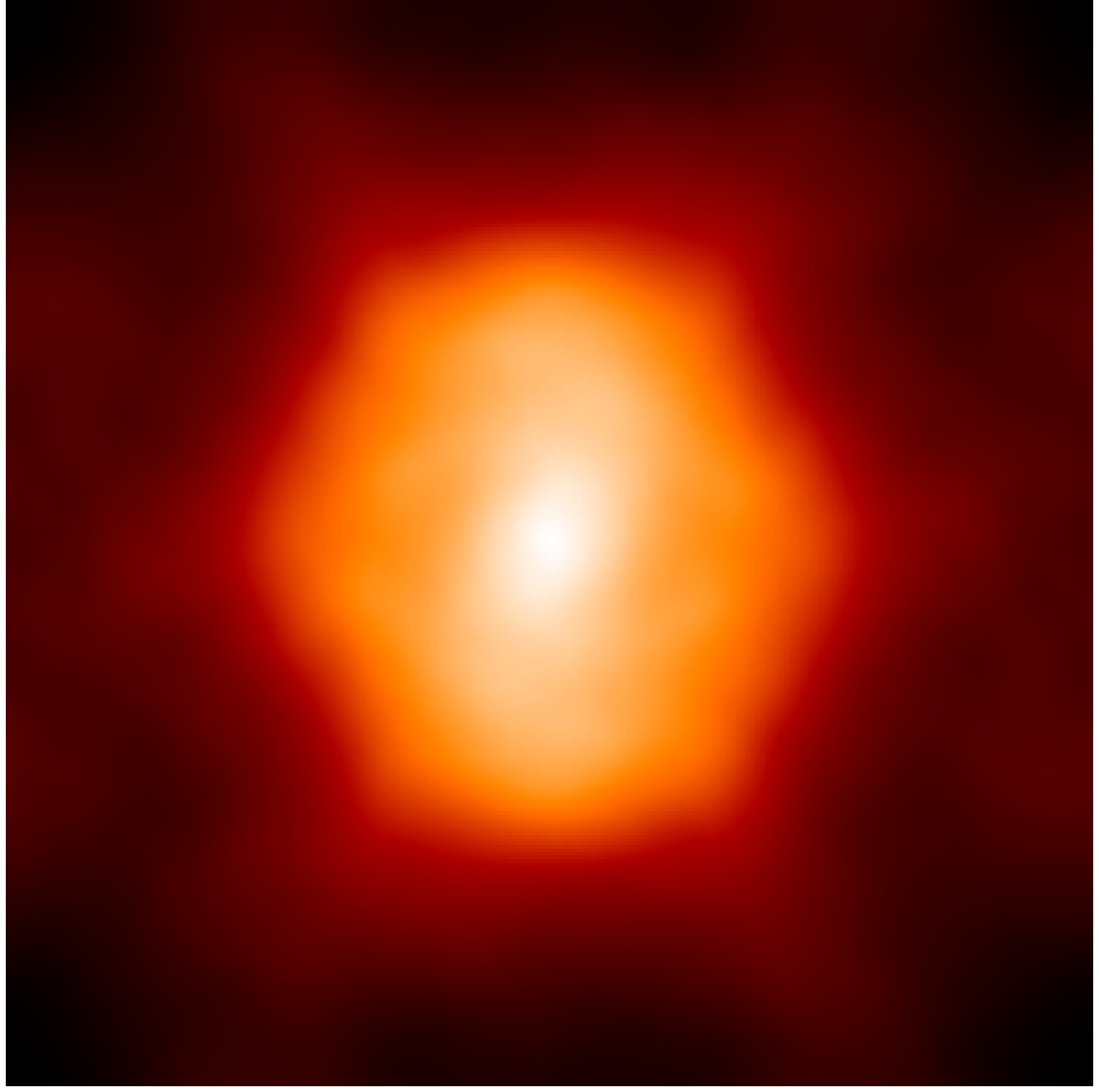,angle=0,width=6cm}}
\caption{A spherical hot spot which has been deconstructed at
different levels. On the top left hand panel we have the pure 
non-Gaussian
signal. The angles $\theta_i$ have been
redrawn uniformly on the top right picture. On the bottom left
the phases $\phi_i$ were redrawn unformly. On the bottom right
we applied an independent unformly distributed  rotation on 
all rings in Fourier space.  From top to
bottom and left to right, a plain regular sphere, a shapeless
sphere, a delocalized sphere, and a disconnected sphere.}
\label{fermagsph}
\end{figure}

\begin{figure}
\centerline{\epsfig{file=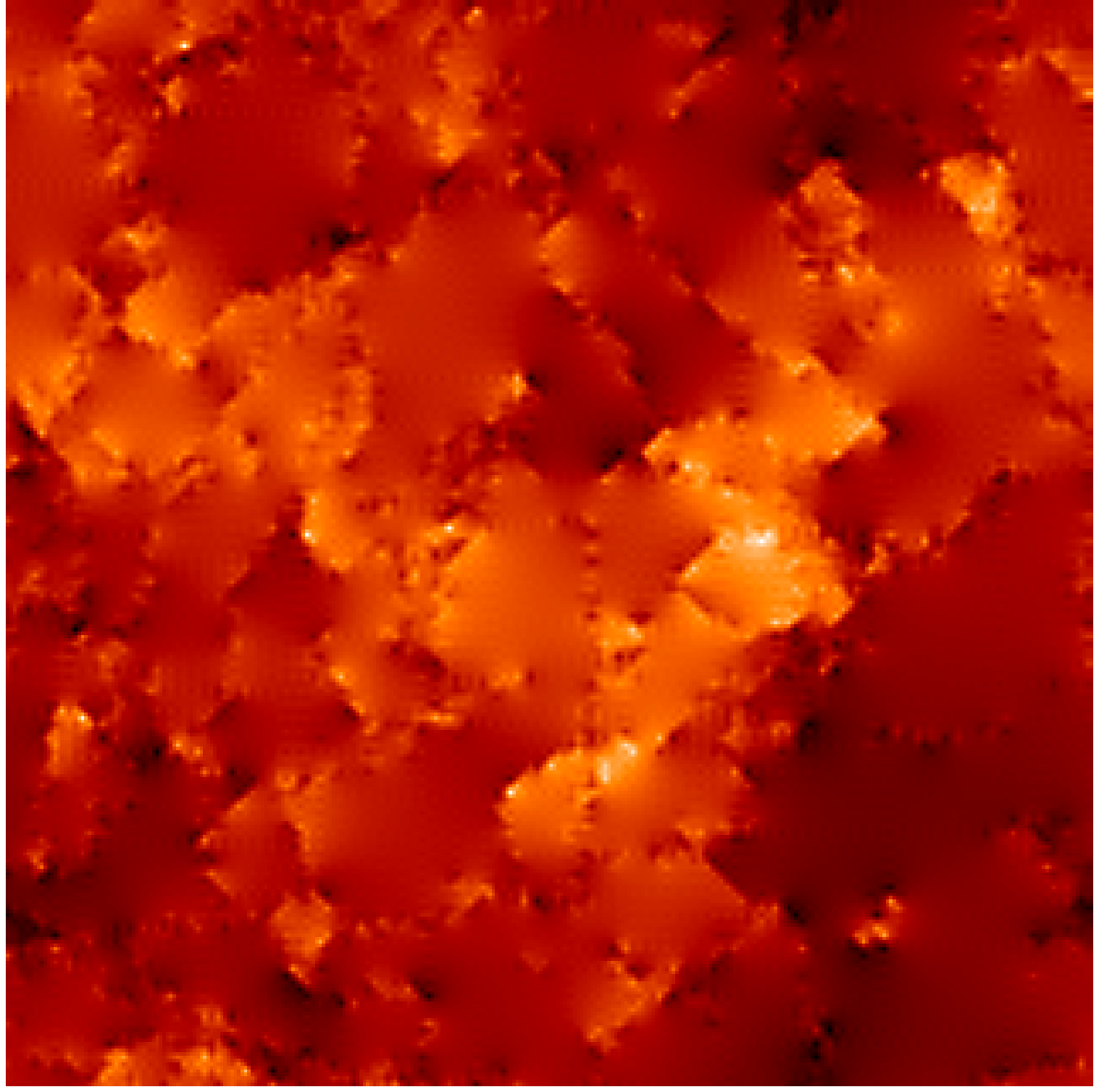,angle=0,width=6cm}
\epsfig{file=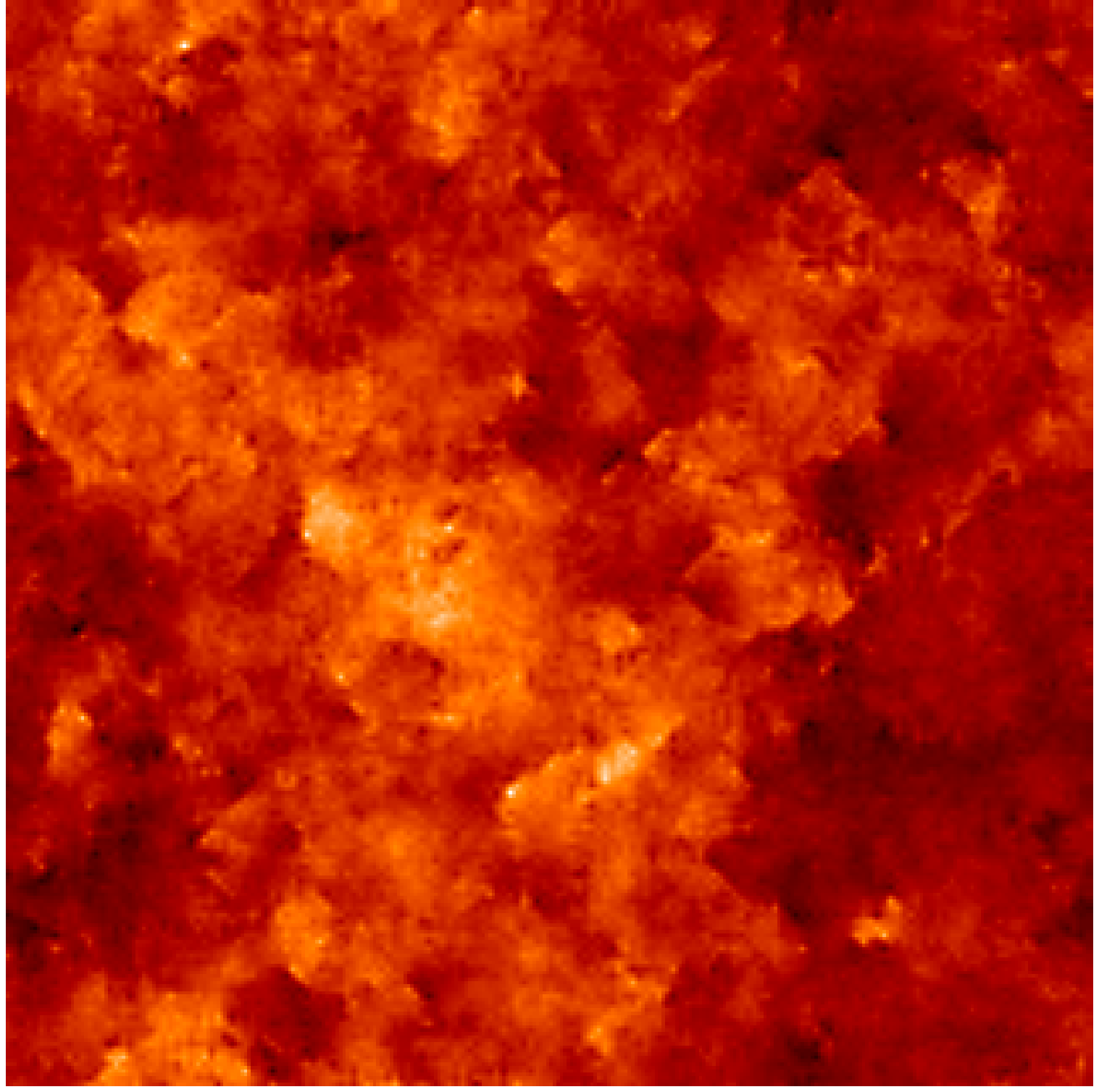,angle=0,width=6cm}}
\centerline{\epsfig{file=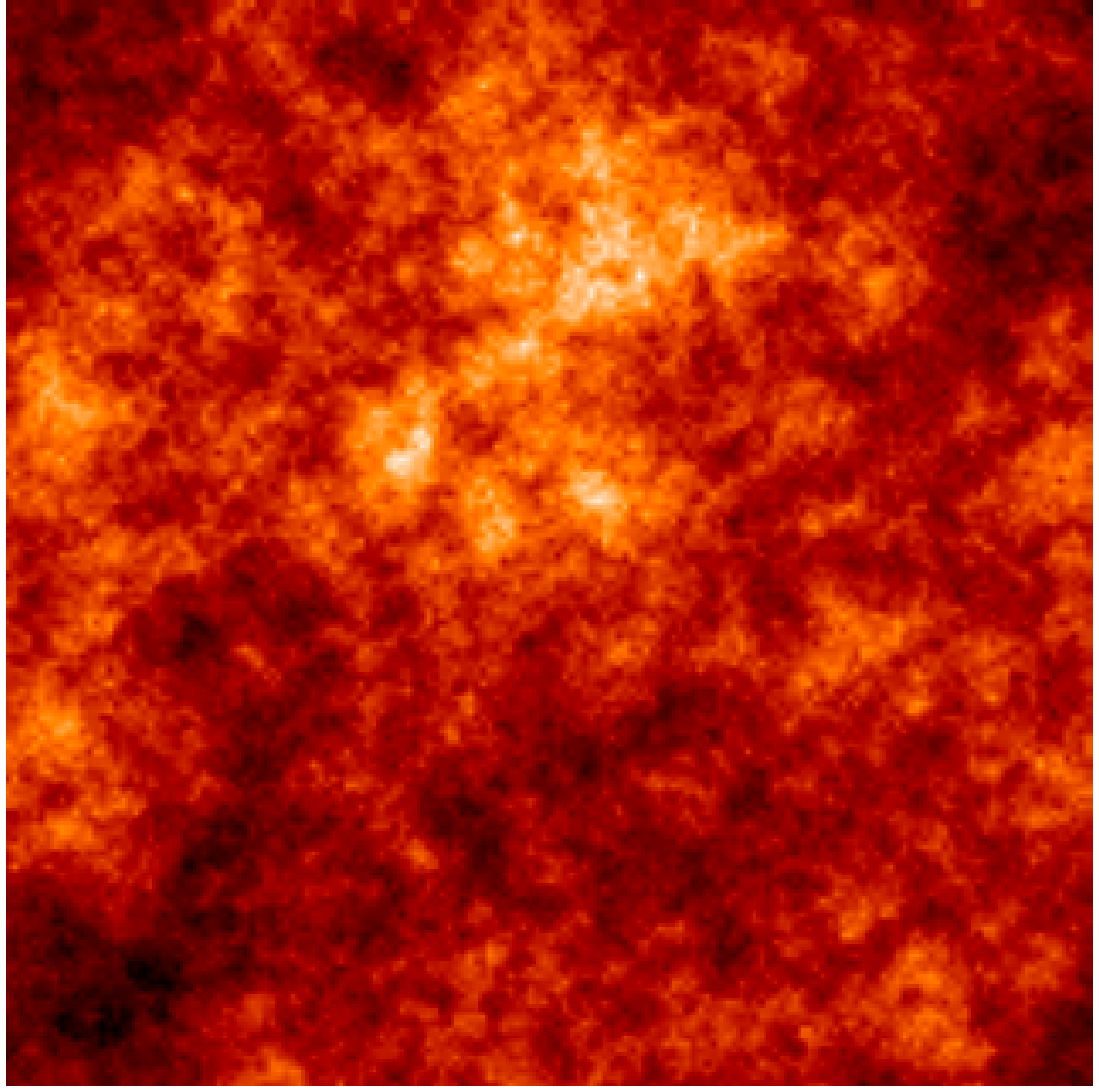,angle=0,width=6cm}
\epsfig{file=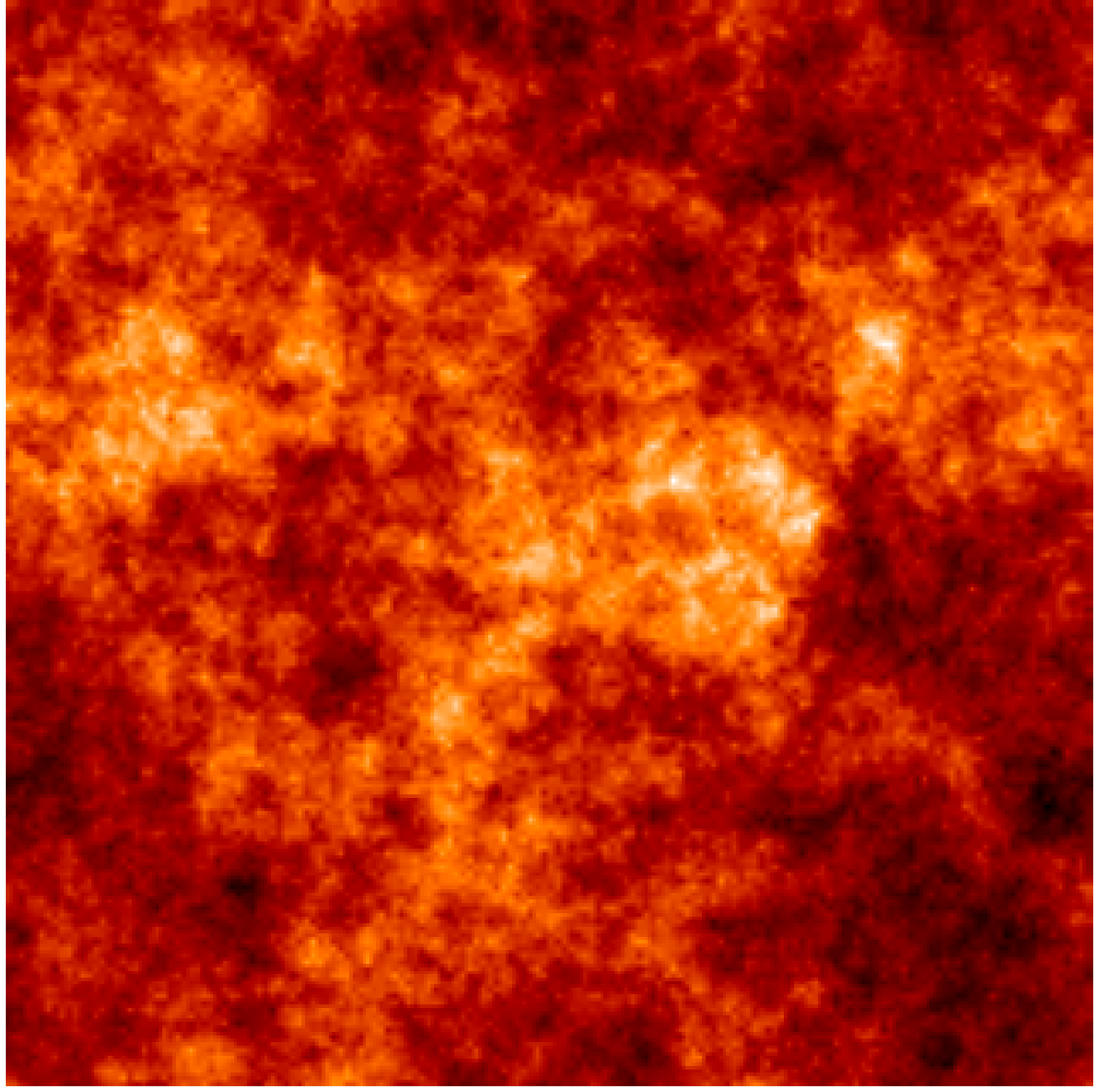,angle=0,width=6cm}}
\caption{The Kaiser-Stebbins effect (top left) and its various
stages of deconstruction. The angles $\theta_i$ have been
redrawn uniformly on the top right picture. On the bottom left
the phases $\phi_i$ were redrawn unformly. On the bottom right
we applied an independent unformly distributed  rotation on 
all rings in Fourier space.  Respectively we have strings,
shapeless strings, unlocalized strings and disconnected strings. }
\label{fermagst}
\end{figure}
The decomposition $\{C(k),{\theta},\phi,\psi\}$ has 
an immediate physical interpretation. The angles $\theta$
reflect the angular distribution of power, and therefore reflect
shape. The phases $\phi$ transform under translations and so contain
the information on position and localization of the structures in the field.
The angles $\psi$ correlate different scales, and therefore tell us
how connected the structures are. For a Gaussian random field
the variables $\{{\theta},\phi,\psi\}$ are all uniformly
distributed reflecting complete lack of structure besides the power spectrum.
In terms of the various levels of structure considered we
can then characterize Gaussian fluctuations as shapeless, delocalized
and disconnected. By comparison with a Gaussian we may then define
structure at different levels. We will say that fluctuations for which
${\theta}$ are not uniformly distributed are shapeful.
If the $\phi$ are not uniformly distributed we shall say the
fluctuations are localized. If the $\psi$ are not uniformly distributed 
the fluctuations are connected. Although visual structure has
room within these definitions, they are considerably
more abstract and general. We may consider highly non visual
types of structure such as shapeful but delocalized fluctuations
or disconnected localized stringy fluctuations. In this sense
we regard our formalism as a robust definition of structure, which goes 
beyond what is visually recognazible and so is tied down to 
our particular and narrow path of natural selection. We may
imagine an alien civilization with Fourier space eyes (say
interferometric eyes \cite{nota}), and a brain trained to recognize
Fourier space structure at many different levels, 
structure that would seem totally non obvious
to our human eyes.

To illustrate the limitations of human vision
we shall now destroy highly structured 
maps level by level, that is Gaussianize only one of the
variable types $\{{\theta},\phi,\psi\}$. Initially there
will be structure at every level, shape, position, and connectedness.
We will remove structure gradually, a fact not disasterous
for the alien civilization referred above, but which will
illustrate the  limitations of the human visual method for recognizing
non-Gaussianity. In Figure~\ref{fermagsph} we play this
game with a sphere. We depict a spherical hot spot in real space, then
a shapeless sphere, a delocalized sphere, and a disconnected 
sphere.  For the case of a sphere we find that what we recognize as shape
is mostly localization. A shapeless sphere keeps its 
recognizable features. On the other hand a delocalized sphere
loses it characteristic features. Indeed the idea of a shapeful but
non-localized object sounds somewhat surreal for all we can 
visually conceptualize. Nevertheless our formalism
will acuse the strong but not obvious non-Gaussianity exhibited
by a delocalized sphere. 

In Figure~\ref{fermagst} we repeat the same
exercise for a map displaying the Kaiser-Stebbins effect from
cosmic strings. Shapeless strings, delocalized strings, and disconnected
strings are shown. Considerable disarray is introduced
in every case, but one may say that disconnected strings as well as
delocalized strings are 
perhaps the most messy of them. This is consistent with the strong signal
in $\psi$ we have found for the case of the realistic Kaiser Stebbins effect.
On the other hand the fact that line-like discontinuities are present 
even for shapeless strings shows how much more structure
there is in the map on top of the structure
which we can recognize. This is important since the beautiful patchwork
is very fragile to the hard realities of noise
and  supperposed Gaussian signal.
In the real world, it turns out, the non-visual feature which
is the connectedness of strings happens to survive much better than the
patchwork (which reflects mostly localization).

\section{Non-Gaussian spectra and cosmic string detection}
\begin{figure}
\centerline{\epsfig{file=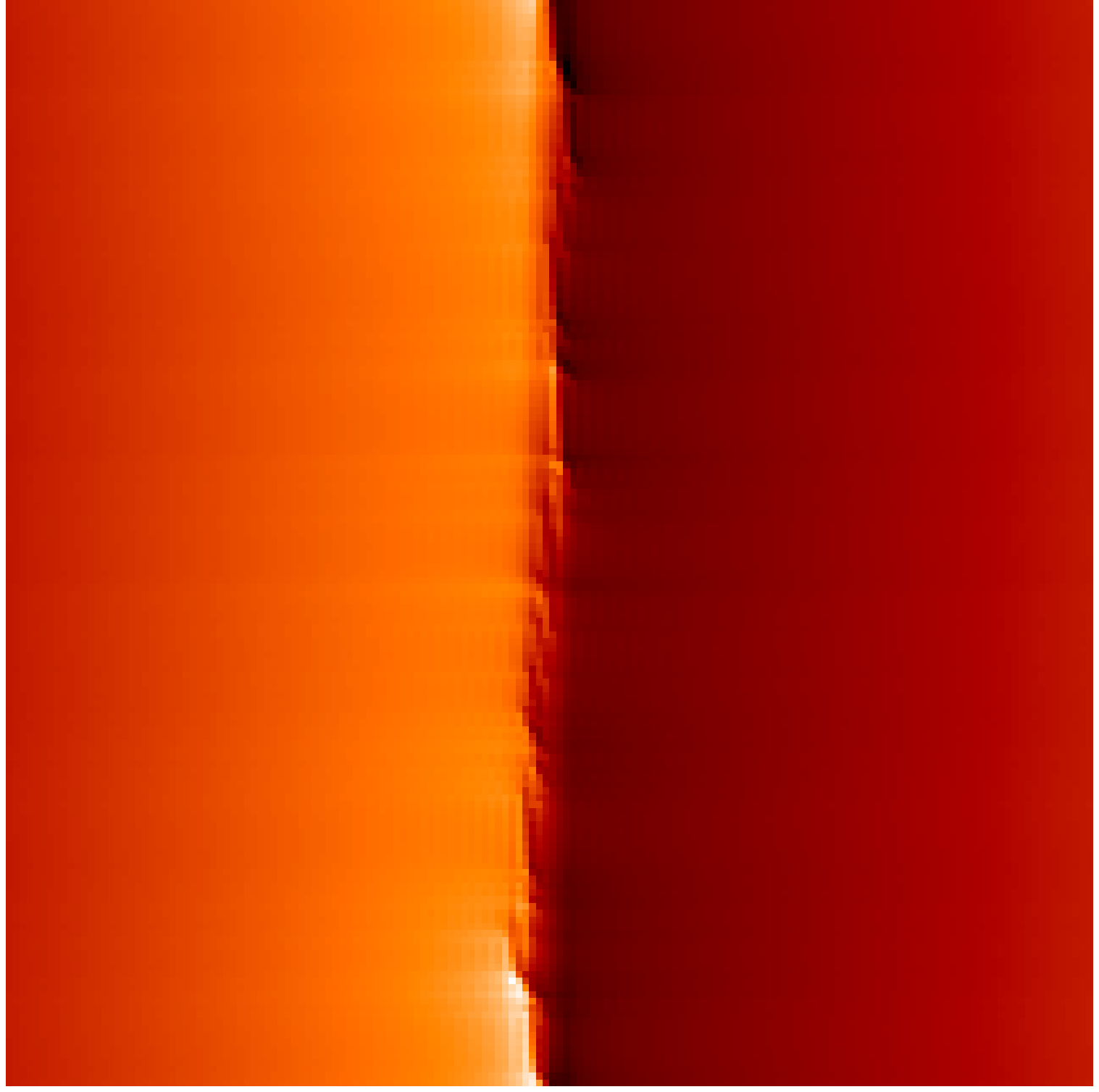,angle=0,width=6cm}
\epsfig{file=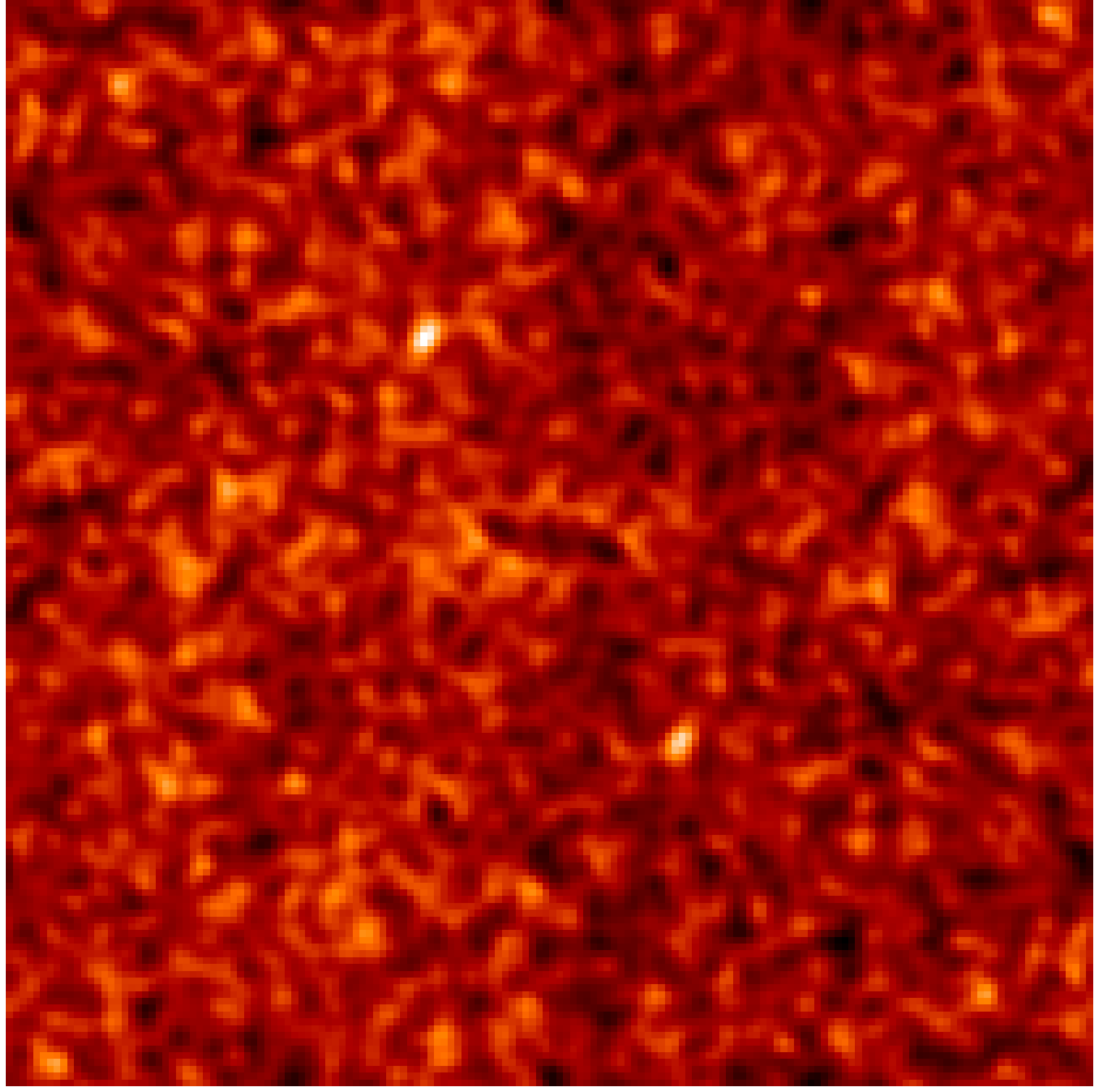,angle=0,width=6cm}}
\caption{A field of the order of $20'$ across showing the Kaiser-Stebbins
effect from a cosmic string before and after the Gaussian signal is added.
}
\label{cs}
\end{figure}

As shown in Sec.3 the existence of a Gaussian background in 
string scenarios implies the need for higher resolution than
thought before in order to detect cosmic string non-Gaussianity. 
Interferometer experiments are therefore favoured.
For such high resolution interferometers can only look into small fields.
We therefore concentrate on a field of some $20'$ across. 
The Kaiser-Stebbins effect before and after the addition of
the Gaussian signal would then look like Fig.~\ref{cs}.

\begin{figure}
\centerline{\epsfig{file=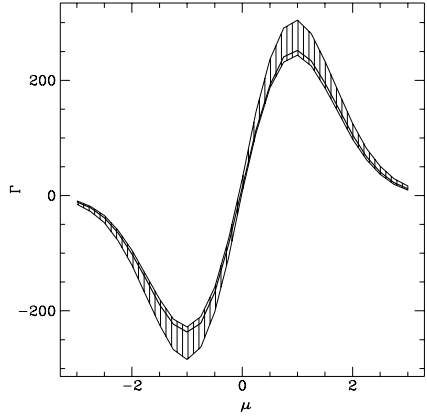,angle=0,width=6cm}}
\caption{The mean Euler characteristic, $\Gamma$ as a function of
threshold for a string map with Gaussian noise (solid line)
and for a pure Gaussian map with the same power spectrum
(the shaded region is the 1 $\sigma$ region around the Gaussian mean,
estimated from a 100 realizations).
}
\label{gen}
\end{figure}

Clearly the string is no longer visible. What is more, we have checked that
the traditional non-Gaussianity tests fail to detect the hidden
small scale non-Gaussianity in the right picture in Fig.~\ref{cs}. 
We have checked this
statement with: pixel histograms, skewness and kurtosis, density of
peaks, topological tests, the 3-point correlation function, and other
traditional tests. As an example we show in Fig.~\ref{gen} the failure
of the Euler characteristic test to detect the realistic Kaiser-Stebbins
effect.

\begin{figure}
\centerline{\epsfig{file=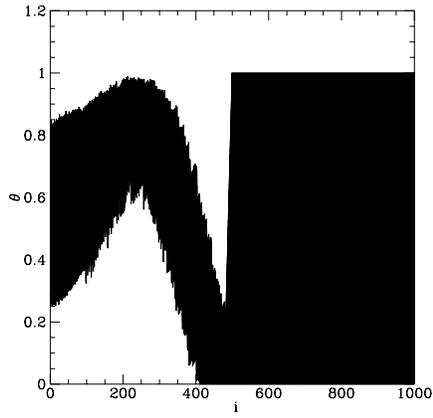,angle=0,width=6cm}}
\caption{The ring spectrum at high $k$ inferred from 100 realizations.
The shaded region
represents a probability larger than $1 \over e$ for the
the values of $\theta_i$ to occur (this generalises the concept of
1 sigma error bar for non-Gaussian distributions.}
\label{stshape}
\end{figure}

However we found that non-Gaussian spectra may still detect the hidden
string after Gaussian signal has been added. Let us first look at 
shape spectra (Fig.~\ref{stshape}).  
For rings at low $k$ these are very Gaussian. 
However, as we go out into high $k$ rings, something
peculiar starts to happen. The straight string spectrum is of course
never visible. However a clear ridge in the distribution of shapes emerges.
This corresponds to the wiggly string shape. There is naturally cosmic/sample
variance in the shape spectrum. This is because not all wiggly strings have
the same wiggles. Therefore, as in the case of $C_\ell$ spectra,
one must complement the average spectrum with error bars. Still the concept
of shape spectrum makes sense, because shapes are not uniformly distributed,
but rather have a peaked distribution.

\begin{figure}
\centerline{\epsfig{file=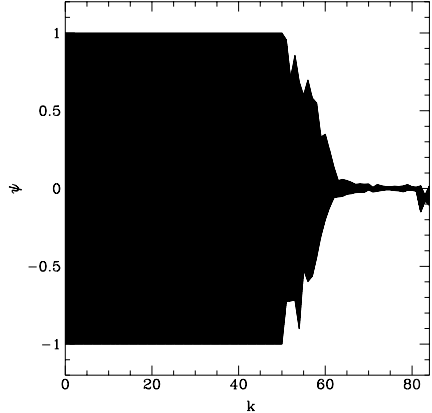,angle=0,width=6cm}}
\caption{The inter-ring spectrum $\psi$.
 The shaded region
represents a probability larger than $1 \over e$ for the
the values of $\psi_i$ to occur.}
\label{stpsi}
\end{figure}

More impressive still is the inter-ring spectrum, shown in Fig.~\ref{stpsi}. 
If an object is exactly 
stringy then it concentrates the power in Fourier space along one
direction. More importantly, this direction is the same for adjacent 
rings. Then, for wiggly strings, one may expect that the direction 
where the maximum modulus is achieved is
highly correlated from ring to ring. As we see in Fig.~\ref{stpsi}, 
as we go up in
$k$ the angle $\psi$ between maximal directions is at first uniformly
distributed. Gradually, as the string signal starts to dominate,
the $\psi$ distribution becomes highly peaked around $\psi=0$. This is a
very impressive signal, with a very small error bar. We put most
of our hopes of detection on this signal.

\section{Prospects for the future}
The statistics presented here should be comprehensive in their
detection of non-Gaussianity, but naturally we are concentrating on
finding characteristic signatures for string
networks  in this framework. 

\subsection{Simulating the sky}
We plan to test our statistics on actual observations. But before 
using real data, we need to test the method on simulations.  We
are currently working on simulated brownian strings.  These are useful
in testing our intuition about the statistics but of course are not
the most realistic examples of strings. The most accurate string 
maps are generated by evolving a string network
from its initial conditions up to the present. We will do this
by means of the integer string code described in \cite{coulson}. 
We hope to generate 10000 maps in this way.    

Once we have a string map, the photon temperature
map is calculated using Laplace's equation \cite{laplace}
\begin{eqnarray}
{\nabla}^2 \frac{\Delta T}{T}=-8\pi G{\vec \nabla}\cdot {\vec U}
\end{eqnarray} 
where all operations here are on the plane orthogonal to some
direction vector ${\vec n}$.
If ${\vec n}$ is chosen to be in the $z$ direction then we have
\begin{eqnarray}
U_i=-\int_{-\infty}^{+\infty}
[\Theta_{0i}(t,X_\gamma(t))-\Theta_{i3}(t,X_\gamma(t))]dt
\end{eqnarray}
where $\Theta_{\mu\nu}$ is the stress-energy tensor for a string and
$i=1,2$. Laplace's  equation is most easily solved in Fourier space so
we are like an interferometer experiment, which observes directly in
Fourier space.

We will assume the fluctuations on the last scattering surface to be
Gaussian, so all we need to specify is the power spectrum.  As the
height of the peak in the spectrum is unknown for defect models and
its position is uncertain we need to fit parameters to the $C_\ell$.

\subsection{Simulating the experiment}
First we have to take into account effects of limited sky coverage and
resolution.  This is done for an interforemeter by modelling the
primary beam of the antennae with a window function $W(\bf{x})$ in real
space and the synthesized beam by a function $B(\bf{x})$.  The Fourier
transform of $B(\bf{x})$, $\tilde{B}(\bf{k})$, is 1 where the Fourier
domain is sampled and 0 otherwise so it is effectively a window in
Fourier space.  Therefore the limit of resolution in real space places
an upper limit on the values of $k$ we can look at.  Large values of $k$
are precisely what we want to look at, so we hope for a high
resolution experiment. The Fourier transform of 
$W(\bf{x})$, $\tilde{W}(\bf{k})$,
corresponds to a limit in the resolution in Fourier space since 
$W(\bf{x})$ cuts
off large scales in real space. This introduces correlations in the
$a(\bf{k})$, so we must be careful in deciding how to sample Fourier space.

Foreground emission can be split into two main components: smooth emission
from dust, synchrotron and free-free emission and point
sources. Smooth emission gives approximately constant errors for all
$k$ so it can be modelled as Gaussian noise (see below).  Spherical
point sources have $\rho_i={\rm constant}$ in each ring, so they should
not interfere with the ring spectra of strings.

Finally we need to add noise to our sky map.  This is assumed to be
Gaussian. It adds an extra term to the covariance matrix of the Fourier modes
$a(\bf k)$ which may be written as \cite{noise}
\begin{equation}
{\langle a^N({\bf k}_i),a^N({\bf
k}_j)\rangle}=\delta_{ij}\frac{\omega^{-1}n_{\rm f}\Omega^3}{(2\pi)^2}
\end{equation}
with $\omega^{-1}=\frac{As^2}{t_{\rm tot}}$,
where $A$ is the area of Fourier space covered, $s^2$ is the sensitivity
of the detector, 
$t_{\rm tot}$ is the total observing time, $n_{\rm f}$ is the number of
fields observed and $\Omega$ is the area of each field.
We will experiment with the various parameters to find the 
best observing strategy.

\subsection{The statistics}
Once we have our maps of stringy skies we can
test the statistics on them so we will be in a position to decide what
the best observing strategies are ({\it ie} how long should be spent in each
field, how many fields should be observed etc).

Other possibilities for statistics to investigate include:
\begin{itemize}
\item
Quantities made up of the phases of the $a(\bf k)$ (in the same way
that the shape spectra are made up of the moduli). The phases contain
all the information about position in real space and so are
complementary to the moduli.
\item
Looking at the two point function around a ring in Fourier space.
\item
Taking another Fourier transform in angle around a ring in Fourier
space.  This and the two point function are alternative ways of
looking at
the distribution of power in rings.
\end{itemize}

Eventually we hope to test our statistics on data from the VLA and
BIMA experiments. These are interferometer experiment with a field of 
a few arcminutes but very high resolution and sensitivity.

\section*{Acknowledgments}
We would like to thank Andy Albrecht, Pedro Ferreira and Shaul Hanany
for discussion. Jo\~ao Magueijo thanks Francesco 
Melchiorri and Monique Signore for organizing such a pleasant workshop.
Finaly we thank PPARC (A.L.) and the Royal Society
(J.M.) for financial support.

\end{document}